\documentclass{ws-p8-50x6-00}

\begin{document}

\thispagestyle{empty}

\begin{flushright}
hep-ph/0001324
\end{flushright}

\vspace{2.0cm}

\begin{center}
\Large\bf Model-independent determination of $|V_{ub}|$
\vspace*{0.3truecm}
\end{center}

\vspace{1.8cm}

\begin{center}
\large Changhao Jin\\
{\sl School of Physics, University of Melbourne
\\
Victoria 3010, Australia\\[3pt]
E-mail: {\tt jin@physics.unimelb.edu.au
         }}
\end{center}

\vspace{1.5cm}

\begin{center}
{\bf Abstract}\\[0.3cm]
\parbox{13cm}
{
The decay distribution of the kinematic variable $\xi_u$ in
inclusive charmless semileptonic decays of $B$ mesons is unique.
The novel method for a model-independent determination of
$|V_{ub}|$ is described.
}
\end{center}

\vspace{2.5cm}

\begin{center}
{\sl To appear in Proceedings of  
the 3rd International Conference on B Physics and CP Violation\\ 
Taipei, Taiwan, December 3-7, 1999}
\end{center}

\newpage
\setcounter{page}{1}
\thispagestyle{empty}


\title{Model-independent Determination of $|V_{\lowercase{ub}}|$}

\author{Changhao Jin}

\address{School of Physics, University of Melbourne, Victoria 3010,
Australia\\E-mail: jin@physics.unimelb.edu.au}


\maketitle

\abstracts{
The decay distribution of the kinematic variable $\xi_u$ in
inclusive charmless semileptonic decays of $B$ mesons is unique.
The novel method for a model-independent determination of
$|V_{ub}|$ is described.
}

\section{Problems and Solutions}
The fundamental Cabibbo-Kobayashi-Maskawa (CKM) matrix element 
$|V_{ub}|$ has been determined from inclusive charmless semileptonic 
decays of $B$ mesons, $B\to X_u\ell\nu$. 
However, there are experimental and theoretical
problems that obstruct a precise determination of $|V_{ub}|$.
Experimentally, it is very difficult to separate signals from
the rare $B\to X_u\ell\nu$ decay from large $B\to X_c\ell\nu$
background. Theoretically, QCD uncertainties arise in calculations
that relate the measured quantity to $|V_{ub}|$.
The potential theoretical uncertainties from 
perturbative and nonperturbative QCD can be comparable.

The solutions for the problems are provided by a novel 
method.\cite{new,ana}
It has been proposed to use the kinematic cut on the 
variable $\xi_u= (q^0+|{\bf q}|)/M_B$ ($q$ is the momentum transfer
to the lepton pair) above the kinematic limit
for $B\to X_c\ell\nu$, $\xi_u>1-M_D/M_B$, to separate 
$B\to X_u\ell\nu$ signal from $B\to X_c\ell\nu$ background.
Most of $B\to X_u\ell\nu$ events
pass the above cut. This kinematic requirement provides a very
efficient way for background suppression. $|V_{ub}|$ can then be
extracted from the weighted integral of the measured $\xi_u$
spectrum via the sum rule for inclusive charmless semileptonic decays
of $B$ mesons with little theoretical uncertainty. 
The sum rule is derived from the light-cone expansion 
and beauty quantum number conservation. Thus a model-independent
determination of $|V_{ub}|$ can be achieved, minimizing the overall
(experimental and theoretical) error. 

\section{Sum Rule}
Because of the large $B$ meson mass, the light-cone expansion is applicable 
to inclusive $B$ decays that are dominated by light-cone singularities.
For inclusive charmless semileptonic decays of $B$ mesons,  
the light-cone expansion and beauty quantum number conservation lead
to the sum rule \cite{new}
\begin{equation}
S\equiv\int_0^1 d\xi_u\, \frac{1}{\xi_u^5}
\frac{d\Gamma}{d\xi_u}(B\to X_u\ell\nu) 
= |V_{ub}|^2\frac{G_F^2M_B^5}{192\pi^3}\, .
\label{eq:sumrule}
\end{equation}
This sum rule has the following advantages:
\begin{itemize}
\item Independent of phenomenological models
\item No perturbative QCD uncertainty
\item Dominant hadronic uncertainty avoided
\end{itemize}
The sum rule (\ref{eq:sumrule}) establishes a relationship between
$|V_{ub}|$ and the observable quantity $S$ in the leading twist approximation 
of QCD. The only remaining theoretical uncertainty in the relation comes from
higer-twist corrections to the sum rule, which are suppressed by a power of 
$\Lambda^2_{\rm QCD}/M_B^2$. 

\section{The $\xi_{\lowercase{u}}$ Spectrum}
Now let me explain why the decay distribution of the kinematic variable $\xi_u$
is unique and why the kinematic cut on $\xi_u$ is very efficient in the 
discrimination between $B\to X_u\ell\nu$ signal and $B\to X_c\ell\nu$ 
background.

Without QCD corrections, the tree-level $\xi_u$ spectrum in the 
free quark decay $b\to u\ell\nu$ in the $b$-quark rest frame 
is a discrete line at $\xi_u= m_b/M_B$. This is 
simply a consequence of kinematics that fixes $\xi_u$ to the single value
$m_b/M_B$, no other values of $\xi_u$ are kinematically allowed in  
$b\to u\ell\nu$ decays. This discrete line at $\xi_u= m_b/M_B \approx 0.9$
lies well above the charm threshold, $\xi_u> 1-M_D/M_B = 0.65$.

The $O(\alpha_s)$ perturbative QCD correction to the $\xi_u$ spectrum
has been calculated.\cite{ana} 
The $\xi_u$ spectrum remains a discrete line at $\xi_u = m_b/M_B$, even if
virtual gluon emission occurs. Gluon bremsstrahlung generates a small tail 
below the parton-level endpoint $\xi_u= m_b/M_B$.

To calculate the real physical decay distribution in $B\to X_u\ell\nu$, 
we must also account for hadronic bound-state effects. 
In the framework of the light-cone expansion, the leading nonperturbative
QCD effect is incorporated in the $b$-quark distribution 
function \cite{jin}
\begin{equation}
f(\xi) = \frac{1}{4\pi}\int \frac{d(y\cdot P)}{y\cdot P}\, e^{i\xi y\cdot P}
\langle B|\bar{b}(0)y\!\!\!/{\cal P}exp[ig_s\int_y^0 dz^\mu A_\mu (z)]b(y)
|B\rangle |_{y^2=0}\, ,
\label{eq:def}
\end{equation}
where ${\cal P}$ denotes path ordering. Although several important properties
of it are known\cite{jin} in QCD, the form of the distribution function has
not been completely determined.
The distribution function $f(\xi)$ has a simple physical interpretation:
It is the probability of finding a $b$-quark with momentum $\xi P$ inside 
the $B$ meson with momentum $P$. The real physical spectrum is then obtained 
from a convolution of 
the hard perturbative spectrum with the soft nonperturbative distribution 
function:
\begin{equation}
\frac{d\Gamma}{d\xi_u}(B\to X_u\ell\nu) = \int_{\xi_u}^1 d\xi\, f(\xi)
\frac{d\Gamma}{d\xi_u}(b\to u\ell\nu, p_b=\xi P) ,
\label{eq:convol}
\end{equation}
where the $b$-quark momentum $p_b$ in the perturbative spectrum is replaced by
$\xi P$. The interplay between nonperturbative and perturbative
QCD effects has been accounted for. 

Bound-state effects lead to 
the extension of phase space from the parton level to the hadron level, also
stretch the spectrum downward below $m_b/M_B$, and are solely responsible for
populating the spectrum upward in the gap between the parton-level endpoint
$\xi_u= m_b/M_B$ and the hadron-level endpoint $\xi_u= 1$. 
The interplay between nonperturbative and perturbative QCD effects eliminates
the singularity at the endpoint of the perturbative spectrum, so that the
physical spectrum shows a smooth behaviour over the entire range of $\xi_u$,
$0\leq\xi_u\leq 1$.

Although the monochromatic $\xi_u$ spectrum at tree level is smeared by gluon
bresstrahlung and bound-state effects around $\xi_u= m_b/M_B$, about
$80\%$ of $B\to X_u\ell\nu$ events remain above the charm
threshold. The uniqueness of the decay distribution of the kinematic variable
$\xi_u$ implies that the kinematic cut on $\xi_u$ is very efficient in 
disentangling $B\to X_u\ell\nu$ signal from $B\to X_c\ell\nu$ background. 

\section{How Do You Measure $S$?}
To measure the observable $S$ defined in Eq.~(\ref{eq:sumrule}), one needs to
measure the weighted $\xi_u$ spectrum
$\xi_u^{-5}d\Gamma(B\to X_u\ell\nu)/d\xi_u$, using the kinematic cut
$\xi_u > 1-M_D/M_B$ against $B\to X_c\ell\nu$ background.
$S$ can then be obtained from an extrapolation
of the weighted spectrum measured above the charm threshold to low $\xi_u$. 

While the normalization of the weighted spectrum given by the sum rule
(\ref{eq:sumrule}) does not
depend on the $b$-quark distribution function $f(\xi)$, thus being 
model-independent, the shape of the weighted spectrum does.
The detailed analysis is presented in Ref.~\cite{ana}. Gluon
bremsstrahlung and hadronic bound-state effects strongly affect the shape of
the weighted $\xi_u$ spectrum. However, the shape of the weighted $\xi_u$
spectrum is insensitive to the value of the strong coupling $\alpha_s$, varied
in a reasonable range. The overall picture appears to be that the weighted 
$\xi_u$ spectrum is peaked towards larger values of $\xi_u$ with a narrow 
width. The contribution below $\xi_u = 0.65$ is small and relatively 
insensitive to forms of the distribution function. This suggests that 
extrapolating the weighted $\xi_u$ spectrum down to low $\xi_u$ would not 
introduce a considerable uncertainty in the value of $S$.  

\section{Summary} 
The kinematic cut on $\xi_u$, $\xi_u > 1-M_D/M_B$, and the semileptonic $B$ 
decay sum rule, Eq.~(\ref{eq:sumrule}), make
an outstanding opportunity for the precise determination of $|V_{ub}|$ from
the observable $S$.
This method is both exceptionally clean theoretically and very 
efficient experimentally in background suppression. 

There remain two kinds of theoretical error in the model-independent 
determination of $|V_{ub}|$. First, higher-twist (or power suppressed) 
corrections to the sum rule cause an error of the order 
$O(\Lambda^2_{\rm QCD}/M_B^2)\sim 1\%$ in $|V_{ub}|$.
Second, the extrapolation of the weighted $\xi_u$ spectrum to low $\xi_u$
gives rise to a systematic error in the measurement of $S$. The size of this 
error depends on how well the weighted spectrum can be measured, since
the measured spectrum would directly determine the form of the distribution
function. In addition, the form of the universal distribution function can 
also be 
determined directly by the measurement of the $B\to X_s\gamma$ photon energy 
spectrum.\cite{jin} The experimental determination of the distribution
function would provide a model-independent way to make the extrapolation,
allowing an error reduction. 

Eventually, the error in $|V_{ub}|$ determined by this method would mainly
depend on how well the observable $S$ can be measured. To measure $S$ 
experimentally one needs to be able to reconstruct the neutrino. This poses
a challenge to experiment. The unique potential of determining 
$|V_{ub}|$ warrants a feasibility study for the experiment.
  
\section*{Acknowledgments}
I would like to thank Hai-Yang Cheng and Wei-Shu Hou for 
the very stimulating and enjoyable conference.
This work was supported by the Australian Research Council.

\end{document}